\documentclass[conference,fleqn, letter, 10pt]{IEEEtran}

\usepackage[table]{xcolor}  
\usepackage{url}
\usepackage{mathtools}
\usepackage{multirow}
\usepackage{caption}
\usepackage{tcolorbox}
\usepackage{subcaption}

\usepackage{enumitem}
\usepackage[style=ieee,maxnames=1,minnames=1]{biblatex}
\AtBeginBibliography{\small}
\usepackage{lipsum}
\usepackage{stfloats}

\usepackage{acronym} \acrodef{5G}[5G]{Fifth Generation of Mobile Networks}
\acrodef{QoE}[QoE]{Quality of Experience}
\acrodef{AI}[AI]{Artificial Intelligence}
\acrodef{BS}[BS]{Base Station}
\acrodef{BS}[BS]{Base Station}
\acrodef{CN}[CN]{Core Network}
\acrodef{PN}[PN]{Processing Node}
\acrodef{NFV}[NFV]{Network Function Virtualization}
\acrodef{SDN}[SDN]{Software-Defined Networking}
\acrodef{RAN}[RAN]{Radio Access Network}
\acrodef{vRAN}[vRAN]{Virtualized \ac{RAN}}
\acrodef{MNO}[MNO]{Mobile Network Operator}
\acrodef{CU}[CU]{Central Unit}
\acrodef{DU}[DU]{Distributed Unit}
\acrodef{vCU}[CU]{Virtualized Central Unit}
\acrodef{vDU}[DU]{Virtualized Distributed Unit}
\acrodef{RU}[RU]{Radio Unit}
\newacroindefinite{RU}{an}{a}
\acrodef{O-RAN}[O-RAN]{Open \ac{RAN}}
\acrodef{ILP}[ILP]{Integer Linear Program}
\acrodef{TTI}[TTI]{Transmission Time Interval}
\acrodef{URLLC}[URLLC]{Ultra-Reliable Low-Latency Communications}

\usepackage{geometry}
\usepackage{amssymb}
\usepackage{amsfonts}
\usepackage[ruled, linesnumbered]{algorithm2e}
\usepackage{algpseudocode}
\DeclareMathOperator*{\minimize}{minimize}

\newcommand{\1}{(\textit{i})}
\newcommand{\2}{(\textit{ii})}
\newcommand{\3}{(\textit{iii})}

\usepackage{tikz}
\newcommand{\EmptyCircle}{%
  \tikz[baseline=-0.6ex]\draw[line width=0.4pt] (0,0) circle (0.8ex);%
}

\newcommand{\FullCircle}{%
  \tikz[baseline=-0.6ex]\fill (0,0) circle (0.8ex);%
}

\usepackage{makecell}

\usepackage[table]{xcolor}
\usepackage{color}

\usepackage{units}
\usepackage{soul}

\usepackage{caption}
\title{
 Adaptive Reallocation of RAN Functions for Resilient 6G Networks
}

\makeatletter
\newcommand{\linebreakand}{%
  \end{@IEEEauthorhalign}
  \hfill\mbox{}\par
  \mbox{}\hfill\begin{@IEEEauthorhalign}
}
\makeatother

\author{

	\IEEEauthorblockN{
      Gabriel M. Almeida\IEEEauthorrefmark{1}\IEEEauthorrefmark{2},
      Jacek Kibi\l{}da\IEEEauthorrefmark{2},
      Joao F. Santos\IEEEauthorrefmark{2}, 
      Kleber Vieira Cardoso\IEEEauthorrefmark{1}
    }
	\IEEEauthorblockA{
      \IEEEauthorrefmark{1}\textit{Universidade Federal de Goiás}, Brazil, e-mails: gabrielmatheus@inf.ufg.br, kleber@ufg.br\\
      \IEEEauthorrefmark{2}\textit{Commonwealth Cyber Initiative}, \textit{Virginia Tech}, USA, 	e-mails: \{jkibilda, joaosantos\}@vt.edu	
    }
}

\begin{document}
\bstctlcite{IEEEexample:BSTcontrol}
\makeatletter
\newcommand{\IEEEpubnotice}{%
  \vspace{-3em}
  \noindent
  \begin{minipage}{\textwidth}
    \footnotesize
    This work has been submitted to the IEEE for possible publication.\\
    Copyright may be transferred without notice, after which this version may no longer be accessible.
  \end{minipage}\par
  \vspace{1em}
  \hrule
  \vspace{0.5em}
}
\makeatother

\twocolumn[
\begin{@twocolumnfalse}
\IEEEpubnotice
\maketitle
\end{@twocolumnfalse}
]
\begin{abstract}
    The disaggregation of base stations into discrete \ac{RAN} functions introduces new threats to mobile networks, as failures in one \ac{RAN} function can trigger cascading failures and disrupt the entire functional chain, impacting network performance and leading to outages. In this paper, we propose the first  resilience mechanism  leveraging the adaptive placement of \ac{RAN} functions to mitigate disruptions and recover service continuity in the presence of compromised infrastructure.
    Our model detects disrupted \acp{RU} due to cascading failures, reacts by re-instantiating \acp{CU} and \acp{DU} in alternative cloud locations, and recovers service continuity by reestablishing functional chains.
    We formulate this recovery process as an optimization problem that maximizes post-failure network performance while considering computational and communication constraints of the infrastructure. We numerically evaluated our approach on a real-world mobile network topology under multiple failure scenarios, and demonstrated that our solution recovers up to 70\% higher throughput compared to conventional resilience mechanisms.
\end{abstract}
    
\begin{IEEEkeywords}
Resilience, 6G networks, disaggregated mobile networks, optimization, function reinstantiation
\end{IEEEkeywords}

\acresetall
    
   \section{Introduction}
\label{sec:intro}

A key development towards 6G is the disaggregation of mobile network functionality, decomposing base stations into chains of \acp{CU}, \acp{DU}, and \acp{RU}, each implementing different layers of the \ac{RAN} protocol stack~\cite{santos2025managing}. This modular approach to network management enables mobile carriers to flexibly deploy and scale  \ac{RAN} functions on different cloud locations to achieve distinct objectives, e.g., lowering costs and catering to dynamic demand~\cite{morais:2023}. 
However, disaggregation also increases the susceptibility of mobile networks to outages, as failures in one component can disrupt the function chain and trigger cascading failures~\cite{Xing:2023}. For example, a \ac{DU} failure can disrupt service for all its associated \acp{RU}, whereas a \ac{CU} failure can disrupt service for all its associated \acp{DU} and subsequently, their associated \acp{RU}.

\begin{figure}[t]
  \centering
  \hfill
  \subfloat[a][Operational RAN.]{\includegraphics[width=0.1586\textwidth]{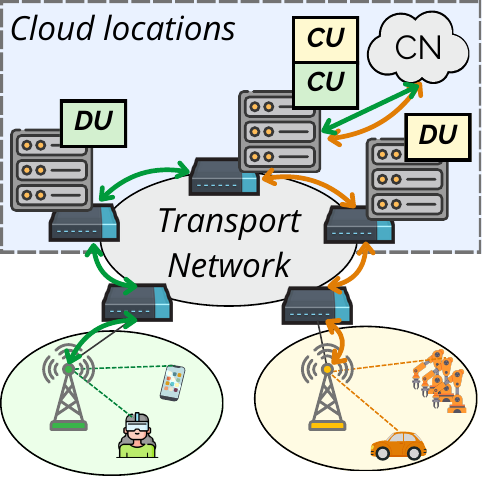}\label{fig:failure_a}}
  \hfill
  \subfloat[b][Failure event.]{\includegraphics[width=0.1586\textwidth]{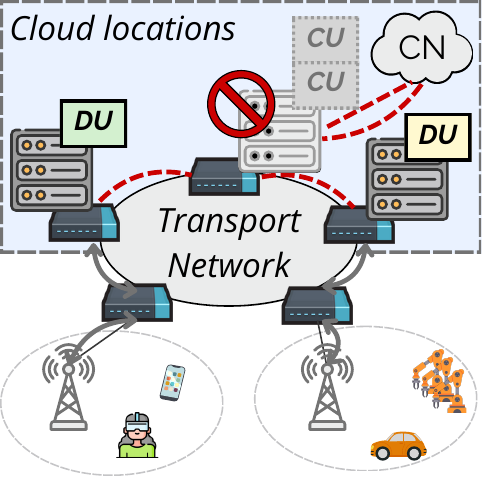} \label{fig:failure_b}}
  \hfill
  \subfloat[c][Adaptive recovery.]{\includegraphics[width=0.1586\textwidth]{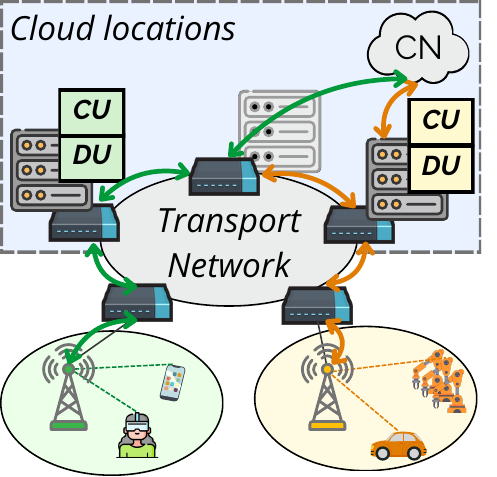}\label{fig:failure_c}}
  \hfill
  \vspace{+1em}
  \caption{Disaggregated mobile networks are susceptible to cascading failures, as disruption in a single \ac{RAN} function can compromise the entire function chain and cause outages.}
  \vspace{-0.8em}
  \label{fig:failrure_scenario}
\end{figure}

At the same time, the softwarized nature of disaggregated mobile networks introduces new opportunities for \textit{resilience} -- the ability to cope with failures and maintain service continuity to minimize the impact from disruptions~\cite{WalidSaad:2024}. 
The softwarization enables the creation of adaptive strategies to respond to disruptions in function chains by re-instantiating \ac{RAN} functions or deploying them in alternative cloud locations to restore service and mitigate the impact of cascading failures~\cite{Lazarev:2023}, as illustrated in Fig.~\ref{fig:failrure_scenario}. While there is a mature body of literature exploring disaggregated mobile network topologies and the optimal placement of \ac{RAN} functions~\cite{morais:2023,Pires:2025,Kim:2025,You:2025,Hojeij:2025}, most of these works focus on improving utility under normal operating conditions, without considering resilience mechanisms to recover from outages and failures.

In this paper, we address this gap and propose the first resilience mechanism for disaggregated mobile networks leveraging the adaptive placement of \ac{RAN} functions. We propose an autonomous solution that \1 detects disrupted \acp{RU} due to cascading failures, \2 reacts by re-instantiating \acp{CU} and \acp{DU} in alternative cloud locations, and \3 recovers service continuity by reestablishing function chains. To the best of our knowledge, this is the first paper to explore the reallocation of \ac{RAN} functions for mitigating function chain disruptions, recovering network operations in the presence of compromised infrastructure. Our contributions are as follows:

\begin{itemize}
\item We propose the first resilience mechanism that adaptively re-instantiates \ac{RAN} functions to restore compromised function chains and recover service to disrupted \acp{RU}.
\item We present a generic mathematical formulation of our proposed resilience mechanism, which jointly captures computing and communication requirements to optimize the recovery of function chains.
\item We validate our proposed resilience mechanism in a real-world mobile network topology, achieving up to $70\%$ performance improvement over existing approaches.
\end{itemize}

The remainder of this paper is organized as follows. In Section~\ref{sec:related_work}, we provide background on robustness, reliability, and resilience, and review related works. In Section~\ref{sec:system_model}, we describe our system model. In Section~\ref{sec:formulation}, we formulate our optimization problem. In Section~\ref{sec:evaluation}, we present numerical results validating our approach. Finally, in Section~\ref{sec:conclusion}, we pose our concluding remarks and directions for future works.

    \section{Technical Background}
\label{sec:related_work}
In this section, we provide technical definitions of resilience, robustness, and reliability. In addition, we examine existing related works on the placement of \ac{RAN} functions, and discuss resilience mechanisms for mobile networks. 

\subsection{Robustness, Reliability, and Resilience}
The concepts of robustness, reliability, and resilience are closely related in the context of mobile networks and collectively contribute to ensuring dependable operations~\cite{Kaada:2022,WalidSaad:2024,Mahmood:2025}. \textit{Robustness} refers to the ability to withstand external disturbances and remain operational, e.g., adapting modulation and coding schemes to counter jamming attacks, while \textit{reliability} denotes the guarantee to deliver the expected service performance, e.g., duplicating packets across base stations to mitigate fading~\cite{Xing:2023}.
\textit{Resilience} pertain the ability to cope with failures 
and recover service continuity aiming to minimize the impact from disruptions
~\cite{Kaada:2022}. 
Only recently, 6G pre-standardization efforts started incorporating resilience as an intrinsic aspect of network design and operation, aiming to create networks that are able to remain operational even under failures by autonomously detecting, reacting, and recovering from disruptions~\cite{nextgtrust, 6gia}.

\subsection{Existing Approaches for the Placement of \ac{RAN} Functions}

The placement of \ac{RAN} functions in disaggregated mobile networks is a joint optimization problem that is concerned with the optimal functional split and the placement of \acp{CU} and \acp{DU} across different cloud locations, creating function chains with the physical \acp{RU} in the network~\cite{morais:2023}. There are numerous works that explored different techniques to optimize the placement of RAN functions. For example, \cite{Pires:2025}~proposed mathematical models to maximize energy efficiency by allocating \acp{CU} and \acp{DU} to processing nodes with lower power consumption; \cite{Kim:2025}~explored cost-aware data-driven scaling to minimize operational costs of \ac{CU} and \ac{DU} placement;
\cite{Hojeij:2025}~employed a data-driven model using LSTM to jointly optimize \ac{CU}/\ac{DU} placement and UE association; and \cite{You:2025}~proposed a channel-aware placement while dynamically activating and deactivating \acp{RU} to reduce power consumption. 

The primary focus of the majority of current research efforts is on developing placement strategies that reduce the cost and improve utility under normal operating conditions, without considering how the network infrastructure would react to disruptions to \acp{CU} and \acp{DU} operation. This includes quantifying the impact of disruptions and recovery strategies, as well as designing placement strategies for \ac{RAN} functions during recovery.

\subsection{Resilience Mechanisms for Mobile Networks}

The resilience of mobile networks has received increasing attention in recent years~\cite{WalidSaad:2024}, with works exploring different mechanisms to introduce resilience in 6G. For example, \cite{Scholler:2013}~presented a service blueprint that specifies requirements for resilience and incorporates service replication to ensure availability; \cite{Carlinet:2019}~proposed an optimization problem for recovering from network disruptions by backing up and re-instantiating affected 5G network functions, and \cite{Lazarev:2023}~considered the resilience of the PHY layer, modeling failures in the digital signal processing as wireless signal impairments, and recovery from failures by instantiating redundant \ac{RAN} functions. While these works improve the network's resilience to failures, they rely on creating replicas of affected components, implicitly assuming that all network resources remain available after disruption. However, this assumption may not always hold true since unexpected events can compromise portions of the mobile network infrastructure~\cite{WalidSaad:2024}. Only a handful of works have explored resilience under such scenarios.

\begin{table}[t]
\centering
\resizebox{0.7\linewidth}{!}{%
\begin{tabular}{|c|c|c|c|}
\hline
\rowcolor[HTML]{DAE8FC} 
Ref.                              & \begin{tabular}[c]{@{}c@{}}RAN Function\\ Placement\end{tabular} & \begin{tabular}[c]{@{}c@{}}Resilience\\ Mechanism\end{tabular} & \begin{tabular}[c]{@{}c@{}}Cascading\\ Failures\end{tabular} \\ \hline
\cite{Pires:2025}    & \FullCircle                                   & \EmptyCircle                                    & \EmptyCircle                                  \\
\rowcolor[HTML]{DAE8FC} 
\cite{Kim:2025}      & \FullCircle                                   & \EmptyCircle                                    & \EmptyCircle                                  \\
\cite{Hojeij:2025}   & \FullCircle                                   & \EmptyCircle                                    & \EmptyCircle                                  \\
\rowcolor[HTML]{DAE8FC} 
\cite{You:2025}      & \FullCircle                                   & \EmptyCircle                                    & \EmptyCircle                                  \\
\cite{Scholler:2013} & \EmptyCircle                                  & \FullCircle                                     & \EmptyCircle                                  \\
\rowcolor[HTML]{DAE8FC} 
\cite{Carlinet:2019} & \EmptyCircle                                  & \FullCircle                                     & \EmptyCircle                                  \\
\cite{Lazarev:2023}  & \EmptyCircle                                  & \FullCircle                                     & \EmptyCircle                                  \\
\rowcolor[HTML]{DAE8FC} 
\cite{Xing:2023}     & \EmptyCircle                                  & \FullCircle                                     & \EmptyCircle                                  \\
\cite{KAADA:24}      & \EmptyCircle                                  & \FullCircle                                     & \EmptyCircle                                  \\ \hline
\rowcolor[HTML]{DAE8FC} 
\textbf{This work}                & \textbf{\FullCircle}                          & \textbf{\FullCircle}                            & \textbf{\FullCircle}                          \\ \hline
\end{tabular}
}
\caption{Qualitative evaluation of related works.}
\label{tab:Related_work}
\end{table}

The work of \cite{Xing:2023} proposes a recovery from compromised \acp{DU} by migrating their user traffic to unaffected \acp{DU}, demonstrating the potential of dynamically managing user associations to maintain service continuity. However, this work only considers failures in the \ac{DU} and does not account for the effects of cascading failures between \acp{CU}, \acp{DU}, and \acp{RU}.
Similarly, the work of \cite{KAADA:24}~focuses solely on the \ac{RU} by dynamically adjusting the transmission power and antenna tilt to expand the coverage area of unaffected \acp{RU} to minimize the effects of \ac{RU} outages and restore service to users. 
In this paper, we leverage the adaptive placement of \ac{RAN} functions to mitigate cascading failures caused by \acp{CU} and \acp{DU} interruptions, enabling the reestablishment of function chains by re-instantiating \ac{CU} and \ac{DU} of \acp{RU} whose function chains have been compromised due to disruptions on the network infrastructure. As such, this work proposes the first comprehensive resilience mechanism for disaggregated mobile networks. Table~\ref{tab:Related_work} compares existing literature with our approach, highlighting the novelty and scope of our contributions.

    \section{System Model}
\label{sec:system_model}

We consider an operational disaggregated mobile network composed of cell sites hosting physical \acp{RU} and cloud locations hosting \acp{DU} and \acp{CU}, e.g., edge, regional and central clouds. These components are interconnected through a transport network, forming a distributed cloud computing architecture. Under operating conditions, the \acp{RU} are connected to \acp{DU}, which are subsequently connected to \acp{CU}, establishing function chains. Finally, each chain is connected to \iac{CN} to serve users.

In this paper, we consider the scenario where the mobile network infrastructure detects failures or disruptions through heartbeat protocols~\cite{Xing:2023}. Upon detection of a failure that interrupts the function chain of an \ac{RU}, we propose an adaptive placement resilience mechanism that reacts by re-instantiating the affected \ac{CU} and \ac{DU} functions on available cloud locations, thereby restoring the \acp{RU} function chains and ensuring the proper service continuity.

\begin{figure}[t]
    \centering
    \includegraphics[width=.90\linewidth]{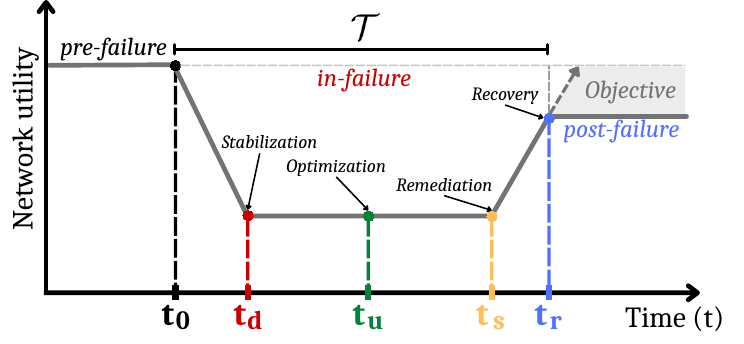}
    \caption{Evolution of the network utility over time, showing the detrimental impact of failures and the objective of our resilience mechanism to restore service and recover utility.}
    \label{fig:utility}
\end{figure}

To capture this dynamic behavior, we discretize time into a set of events $\mathcal{T} = \{t_0, t_d, t_u, t_s, t_r\}$, where each $t \in \mathcal{T}$ represents a significant network state. The instant $t_0$ represents the \textit{pre-failure} state, corresponding to the network's normal operating conditions before any disruption occurs. At time $t_d$, the \textit{in-failure} state begins, the network absorbs the failure's impact on network utility, aiming to stabilize the utility decrease through robustness mechanisms, e.g., user handovers. Then, at $t_u$, the network identifies the failure and triggers our resilience mechanism, at time $t_s$, deciding the optimal recovery solution and initiating the restoration process. Finally, at $t_r$, the \textit{post-failure} state is reached, representing the network's utility after the function chains of the \acp{RU} are re-established and service continuity is recovered. Figure~\ref{fig:utility} illustrates the temporal evolution of our system in terms of network utility.

To model the disaggregated mobile network system, we consider a set $\mathcal{R}$, where each element $r \in \mathcal{R}$ represents an \ac{RU} responsible for providing wireless coverage to users. Each \ac{RU} $r \in \mathcal{R}$ serves a set of users $\mathcal{U}_r$, where each element $u \in \mathcal{U}_r$ denotes a user associated with that \ac{RU}. The cloud computing infrastructure is represented by a set $\mathcal{N}$, where each element $n \in \mathcal{N}$ corresponds to a cloud location with its available capacity at time $t \in \mathcal{T}$ denoted by $\theta_t(n)$. 

Since we consider an operational mobile network, the placement of \acp{CU} and \acp{DU} for all \acp{RU} is known at time $t_0 \in \mathcal{T}$. We denote as $c_{t_0}(n,r) = \{0, 1\}$ the  \ac{CU} placement for \ac{RU} $r \in \mathcal{R}$ at time $t_0 \in \mathcal{T}$ is running at cloud location $n \in \mathcal{N}$. Similarly, we consider as $d_{t_0}(n,r) = \{0, 1\}$ the \ac{DU} placement for \ac{RU} $r \in \mathcal{R}$ at time $t_0 \in \mathcal{T}$. We denote by $\Theta^{CU}_{t}(r)$ and $\Theta^{DU}_{t}(r)$ the computing workloads of the \ac{CU} and \ac{DU} associated with \ac{RU} $r \in \mathcal{R}$ at an arbitrary time $t$, respectively. The throughput requirements for \ac{RU} $r \in \mathcal{R}$ are defined as follows: $\Gamma^{CU}_
{t}(r)$ represents the required throughput between the \ac{CN} and the \ac{CU}; $\Gamma^{DU}_{t}(r)$ corresponds to the throughput between the \ac{CU} and the \ac{DU}; and $\Gamma^{RU}_{t}(r)$ denotes the throughput between the \ac{DU} and the \ac{RU}, at an arbitraty time $t$. Similarly, the latency requirements for these segments are denoted by $\Delta^{CU}_{t}(r)$, $\Delta^{DU}_{t}(r)$, and $\Delta^{RU}_{t}(r)$, respectively.

We represent the topology of the disaggregated mobile network and the cloud computing infrastructure as a graph $\mathcal{G} = (\mathcal{V}, \mathcal{E})$, where $\mathcal{V} = \mathcal{R} \cup \mathcal{N} \cup \{n_0\}$ represents the set of nodes, composed of the \acp{RU}, the cloud locations, and the \ac{CN} denoted by $n_0$. The set of links represents the transport network infrastructure, and is given by $\mathcal{E} = \{(i,j); i, j \in \mathcal{V}\}$, where each link from $i$ to $j$ is represented by the tuple $(i,j)$ with an available bandwidth capacity $\gamma_t(i,j)$ at time $t$. We consider a set of paths $\mathcal{P}_{v,w} $ between $v, w \in \mathcal{V}$ as a sequence of links $(i, j)$, where $i, j \in \mathcal{V}$. $L_p(i, j)$ indicates if link $(i, j)$ is in path $p \in \mathcal{P}_{v,w}$. Each path $p \in \mathcal{P}_{v,w} $ has a corresponding end-to-end latency $\delta_t(p)$ at time $t \in \mathcal{T}$.
    \section{Problem Formulation}
\label{sec:formulation}

In this section, we present the mathematical formulation of our resilience mechanism leveraging the adaptive placement of \ac{RAN} functions. Our objective is to restore network service continuity by re-instantiating \acp{CU} and \acp{DU} of  \acp{RU} whose function chains have been compromised due to disruptions on the disaggregated mobile network infrastructure.

To represent the \textit{in-failure} state of the disaggregated mobile network, we define two subsets of \acp{RU}, $\mathcal{R}^{op}$ and $\mathcal{R}^{dis}$, which represent the sets of operational and disrupted \acp{RU}, respectively, where $\mathcal{R} = \mathcal{R}^{op} \cup \mathcal{R}^{dis}$. Similarly, we denote by $\mathcal{N}^{op}$ the set of cloud locations that remain operational, and by $\mathcal{P}^{op}_{v,w}$ the set of available routing paths between $v, w \in \mathcal{V}$ that are still operational.

To model the re-instantiation of \acp{CU} and \acp{DU}, we define two decision variables, $g_r^n, h_r^n \in \{0,1\}$, indicating whether cloud location $n \in \mathcal{N}^{op}$ will be selected (or not) to re-instantiate the \ac{CU} or \ac{DU} for \ac{RU} $r \in \mathcal{R}^{dis}$, respectively. Similarly, to represent the routing decisions, we define three decision variables, $x_r^p, y_r^p, z_r^p \in \{0,1\}$, indicating whether path $p \in \mathcal{P}^{op}_{v,w}$ will be used (or not) to route the traffic between the (\ac{CN}--\ac{CU}), (\ac{CU}--\ac{DU}), and (\ac{DU}--\ac{RU}) for \ac{RU} $r \in \mathcal{R}^{dis}$, respectively.

We define the optimization objective as minimizing the degradation of network utility resulting from function chain disruptions. Inspired by the general definition of resilience~\cite{WalidSaad:2024}, the goal is to reduce the gap between the utility observed in the \textit{pre-failure} state, denoted by $\mu_{t_0}$, and the expected utility in the \textit{post-failure} state, denoted by $\mu_{t_r}$. The objective function is defined as follows:
\begin{equation}
\label{eq:OF_TP}
    P1: \minimize_{g_r^{n}, h_r^{n}, x_r^p, y_r^p, z_r^p} \left( \mu_{t_0} - \mu_{t_r} \right).
\end{equation}

In this work, we define network utility in terms of the aggregated throughput of all users in the network, providing a direct and measurable indication of resilience~\cite{Kaada:2022}. To capture how effectively the network restores user connectivity after a failure event, we consider the aggregated throughput of operational \acp{RU} and recovered \acp{RU}, based on the re-instantiation of \acp{CU} and \acp{DU}. The utility at an arbitrary time instance $t$ can be expressed as:

\normalsize
\small
\begin{equation}
\label{eq:throughput}
    \mu_t =\,\sum_{r \in \mathcal{R}^{op}} \sum_{u \in \mathcal{U}_r}\,\rho(u, r)\,+\,\sum_{\substack{r \in  \mathcal{R}^{dis}}} \sum_{\substack{m,n \\\in \mathcal{N}^{op}}} \left [ g_r^mh_r^n\sum_{u \in \mathcal{U}_r} \rho(u, r) \right ],
\end{equation}
\normalsize

\noindent where $\rho(u, r)$ denotes the expected throughput of user $u \in \mathcal{U}_r$ connected to \ac{RU} $b \in \mathcal{R}$. 

Our optimization model is also subject to the following constraints. For each \ac{RU} $r \in \mathcal{R}^{dis}$, at most one decision is made for the re-instantiation of its \ac{CU} and \ac{DU} functions:
\begin{equation}
    \sum_{n \in {\mathcal{N}^{op}}} g_r^{n} \leq 1, 
    \qquad \forall \ r \in {\mathcal{R}^{dis}},
\end{equation}
\begin{equation}
    \sum_{n \in {\mathcal{N}^{op}}} h_r^{n} \leq 1, 
    \qquad \forall \ r \in {\mathcal{R}^{dis}}.
\end{equation}

For each \ac{RU} $r \in {\mathcal{R}^{dis}}$, we determine which \ac{RAN} functions need to be re-instantiated and their placement across the available cloud locations. To ensure service availability under the disrupted infrastructure, the processing capacity of each cloud location $n \in {\mathcal{N}^{op}}$ must not be exceeded. Accordingly, the total instantiated workloads must remain within the available computational resources, as enforced by the following:

\normalsize
\footnotesize
\begin{align}
    \sum_{r \in {\mathcal{R}^{dis}}} \bigg ( g_r^n \Theta^{CU}_{t}(r) + h_r^n \Theta^{DU}_{t}(r) \bigg ) \leq \theta_t(n), \ \ \forall n \in {\mathcal{N}^{op}}, \ t \in \mathcal{T}.
\end{align}
\normalsize

In order to reduce the number of migrations, we want to maintain \emph{pre-failure} allocation where no migration is needed:
\begin{equation}
    g_r^n \geq c_{t_0}(n, r), \, \qquad \forall n \in {\mathcal{N}^{op}}, r \in {\mathcal{R}^{dis}},
\end{equation}
\begin{equation}
    h_r^n \geq d_{t_0}(n, r), \qquad \forall n \in {\mathcal{N}^{op}}, r \in {\mathcal{R}^{dis}}.
\end{equation}

Each segment of the function chain must have a routing path $p \in {\mathcal{P}^{op}_{v,w}}$ selected to transmit data between the \ac{CN} and \acp{CU}, between \acp{CU} and \acp{DU}, and between \acp{DU} and \acp{RU}. We enforce this with the following constraints:
\begin{equation}
    \sum_{p \in \mathcal{P}_{n_0, n}^{op}} g^n_r x^p_r = 1, \qquad \qquad \forall n \in {\mathcal{N}^{op}}, r \in {\mathcal{R}^{dis}},
\end{equation}
\begin{equation}
    \sum_{p \in {\mathcal{P}_{m, n}^{op}}} g^m_r h^n_r y^p_r = 1, \qquad \ \ \, \forall m, n \in {\mathcal{N}^{op}}, r \in {\mathcal{R}^{dis}},
\end{equation}
\begin{equation}
    \sum_{p \in {\mathcal{P}_{n, r}^{op}}} h^n_r z^p_r = 1, \qquad \qquad \, \, \forall n \in {\mathcal{N}^{op}}, r \in {\mathcal{R}^{dis}}.
\end{equation}

The capacity constraints of the transport network links $(i, j) \in \mathcal{E}$ must be satisfied to ensure that the total routed traffic does not exceed the available bandwidth. Formally, this requirement is expressed as:
\begin{align}
    \sum_{r \in \mathcal{R}^{dis}}&\sum_{p \in {\mathcal{P}^{op}_{v, w}}} \bigg [ L_{p}(i,j) \bigg ( x^p_r \ \Gamma^{CU}_{t}(r) + y^p_r \ \Gamma^{DU}_{t}(r) \nonumber \\ & + z^p_r \ \Gamma^{RU}_{t}(r) \bigg ) \bigg ] \leq \gamma_t(i, j), \ \ \forall (i, j) \in \mathcal{E}, t \in \mathcal{T}.
\end{align}

The routing paths $p \in {\mathcal{P}_{v,w}}$ selected for each segment of the function chain must satisfy the maximum latency requirements associated with traffic flows between the \ac{CN} and \acp{CU}, between \acp{CU} and \acp{DU}, and between \acp{DU} and \acp{RU}. This requirement is expressed as follows:
\begin{equation}
    \sum_{p \in {\mathcal{P}_{v,w}^{op}}} x^p_r \delta_t(p) \leq \Delta^{CU}_{t}(r), \qquad \forall r \in {\mathcal{R}^{dis}}, t \in \mathcal{T},
\end{equation}\vspace{-0.7em}
\begin{equation}
    \sum_{p \in {\mathcal{P}_{v,w}^{op}}} y^p_r \delta_t(p) \leq \Delta^{DU}_{t}(r), \qquad \forall r \in {\mathcal{R}^{dis}}, t \in \mathcal{T},
\end{equation}\vspace{-0.7em}
\begin{equation}
    \sum_{p \in {\mathcal{P}_{v,w}^{op}}} z^p_r \delta_t(p) \leq \Delta^{RU}_{t}(r), \qquad \forall r \in {\mathcal{R}^{dis}}, t \in \mathcal{T}.
\end{equation}

The P1 objective together with the above constraints form an \ac{ILP} optimization problem~\cite{morais:2023}, which is known to be NP-complete. 
However, as we will show in Section~\ref{sec:evaluation}, this problem for relevant instances of a real-world network topology can be numerically evaluated to optimality under multiple relevant scenarios using CPLEX.

    \section{Numerical Evaluation}
\label{sec:evaluation}

In this section, we evaluate our resilience mechanism based on the adaptive placement of \ac{RAN} functions. We describe our simulation setup and failure scenario. Then, we assess our proposed solution regarding its temporal evolution, recovery performance, throughput resilience, and resource utilization.

\subsection{Simulation Setup and Failure Scenarios}

We developed a simulator to evaluate our solution numerically, leveraging a real-world mobile network topology, obtained from the \textit{5G-Crosshaul} project~\cite{5Gxcrosshaul:17}. The topology comprises 50 \acp{RU}, each co-located with a cloud location, interconnected through a ring-based transport network~\cite{morais:2023}. In our numerical evaluations, we use subsets of this topology, ranging from 5 to 50 \acp{RU}, to evaluate the performance of the recovery mechanisms under different network sizes. Each \ac{RU} operates with a \unit[100]{MHz} bandwidth, 32 antenna ports, 8 MIMO layers, and 256-QAM modulation.
We consider four failure scenarios with increasing severity, in which different percentages of random cloud locations are compromised: low impact (5\%), medium impact (10\%), high impact (25\%), and extreme impact (50\%)~\cite{TRUptime:2024}. These failure scenarios enable us to examine the performance of our resilience mechanism under various conditions. We consider 30 different instances for each scenario and compared the results of our optimization model against three benchmarks: \1 the network operating under normal conditions (\textit{pre-failure} state), \2 the network suffering performance degradation (\textit{in-failure} state), and \3 the recovery mechanism from Kaada~\cite{KAADA:24}, which mitigates failures by adjusting the transmission power and antenna tilt to expand the coverage area of \acp{RU} without re-instantiating \acp{CU} or \acp{DU}. 

\subsection{Responding to Failures and Restoring Service}

First, we are interested in examine the temporal behavior of our resilience mechanism, assessing how it responds to failures and restores service over time. This evaluation allows us to observe the dynamics of utility recovery and qualitatively assess the adaptive behavior of our approach. Figure~\ref{fig:utility_over_time} shows the evolution of network utility over time for our largest topology with 50~\acp{RU}, evaluated across each \ac{TTI} for a \unit[15]{KHz} subcarrier spacing~\cite{TS38211:2025}, under the four different failure scenarios. We can observe how the network utility worsens with the severity of the failure scenarios, demonstrating the impact of compromised cloud locations on service performance. We consider a static wait period of \unit[40]{ms} until recovery is triggered ($t_d$)~\cite{Xing:2023}. We plan on investigating the impact of different periods and adaptive triggers in future work. Once triggered ($t_u$), our optimization model decides the optimal solution to re-instantiate compromised \ac{CU} and \ac{DU} functions ($t_s$). As these new instances become operational, function chains are restored, and the network utility gradually increases until it reaches a new steady state representing the recovered state ($t_r$). The recovery performance naturally depends on the time required to re-instantiate functions, which may vary across different deployment environments.

\iftrue
\begin{figure}[t]
     \vspace{-0.5em}
     \centering
     \includegraphics[width=.99\linewidth]{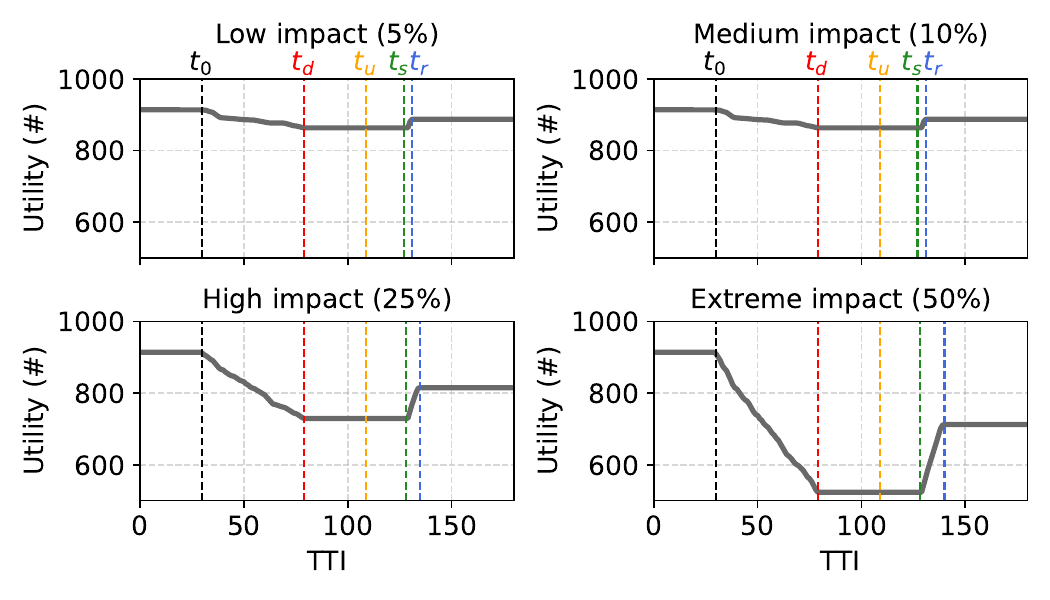}
     \vspace{-2em}
     \caption{Temporal\,evolution\,of\,network\,utility.\,After\,a\,disruption ($t_0$), the\,network\,stabilizes\,($t_d$)\,and\,triggers\,our\,resilience solution ($t_u$)\,defining\,the recovery\,policy\,($t_s$)\,recovering\,utility ($t_r$).}
     \label{fig:utility_over_time}
     \vspace{-0.5em}
\end{figure}

\fi

\subsection{Recovery Performance}

\begin{figure}[ht]
    \vspace{-0.4em}
    \centering
    \includegraphics[width=.99\linewidth]{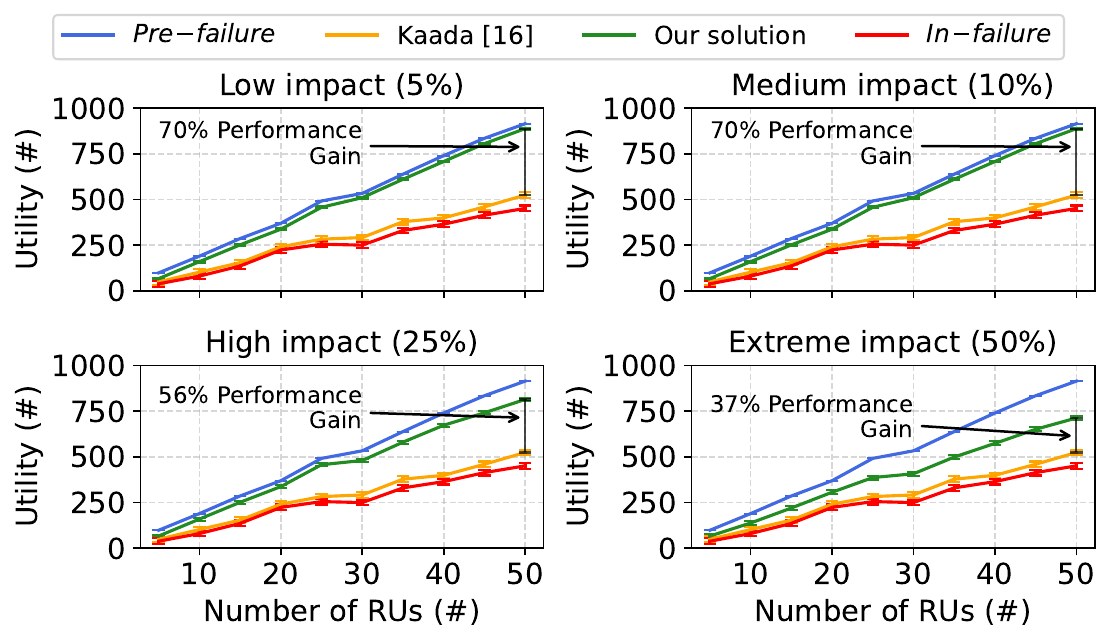}
    \caption{Recovery performance of different resilience mechanisms. Across all scenarios, our solution achieves higher network utility after recovery compared to baseline approaches.}
    \label{fig:utility_comparison}
    \vspace{-0.5em}
\end{figure}

Second, we are interested in assessing the recovery performance of our resilience mechanism, using aggregate user throughput as the network utility metric, as defined in Equation~(\ref{eq:throughput}). 
Figure~\ref{fig:utility_comparison} shows the results of our numerical evaluation for different failure scenarios and network sizes. The \textit{pre-failure} curve represents the network utility when all \acp{RU} operate normally, serving as an upper bound for our evaluation, while the \textit{in-failure} curve represents the network utility when part of the infrastructure is compromised, disrupting function chains and user connectivity, serving as our lower bound. 
We can observe that our solution achieves a 50--83\% performance gain over the baseline network recovery, and 37--70\% gain relative to Kaada~\cite{KAADA:24}. These results show a significant advantage of our approach, which directly responds to failures by re-instantiating \ac{CU} and \ac{DU} to reestablish function chains and restore the operation of \acp{RU} to recover service continuity.

\subsection{Throughput Resilience}

\begin{figure}[ht]
    \vspace{-1em}
    \centering
    \includegraphics[width=.95\linewidth]{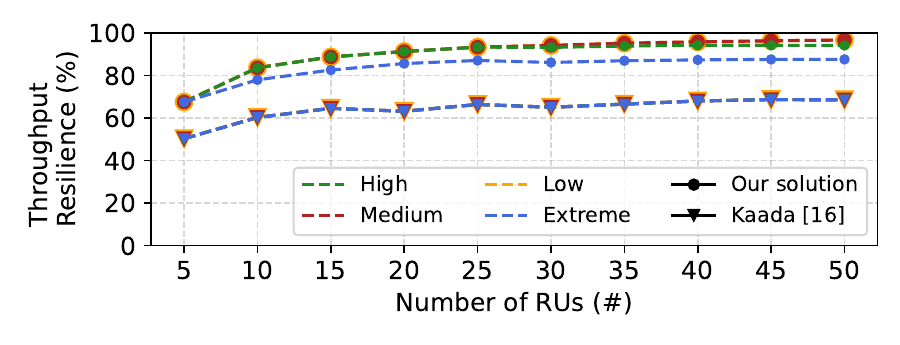}
    \vspace{-0.5em}
    \caption{Comparison of throughput resilience. Our solution achieves higher resilience across all scenarios presenting 80\% recovery for most cases and up to 96\% in larger topologies.}
    \label{fig:resilience_comparison}
    \vspace{-0.5em}
\end{figure}

In this analysis, we are interested in quantifying the resilience of the recovery mechanisms under the different failure scenarios. This assessment is crucial to evaluate their performance and understand how much of the network utility they can recover from varying levels of disruption~\cite{WalidSaad:2024}. Figure~\ref{fig:resilience_comparison} shows the results of our simulations for different failure scenarios and network sizes. We observe that all mechanisms perform better on larger topologies, as these offer more configuration options that can be explored for adapting and recovering from failures. Kaada~\cite{KAADA:24} achieved around 70\% resilience across all failure scenarios, while our solution exceeded 80\% for all instances with more than 15 \acp{RU}. At the size of 50 RUs, our approach reached 96\% throughput resilience, corresponding to only a 4\% service degradation. These results demonstrate the benefits of our mechanism to recover from failures caused by compromised infrastructure.

\subsection{Computational Resource Utilization}

\begin{figure}[t]
    \centering
    \includegraphics[width=.95\linewidth]{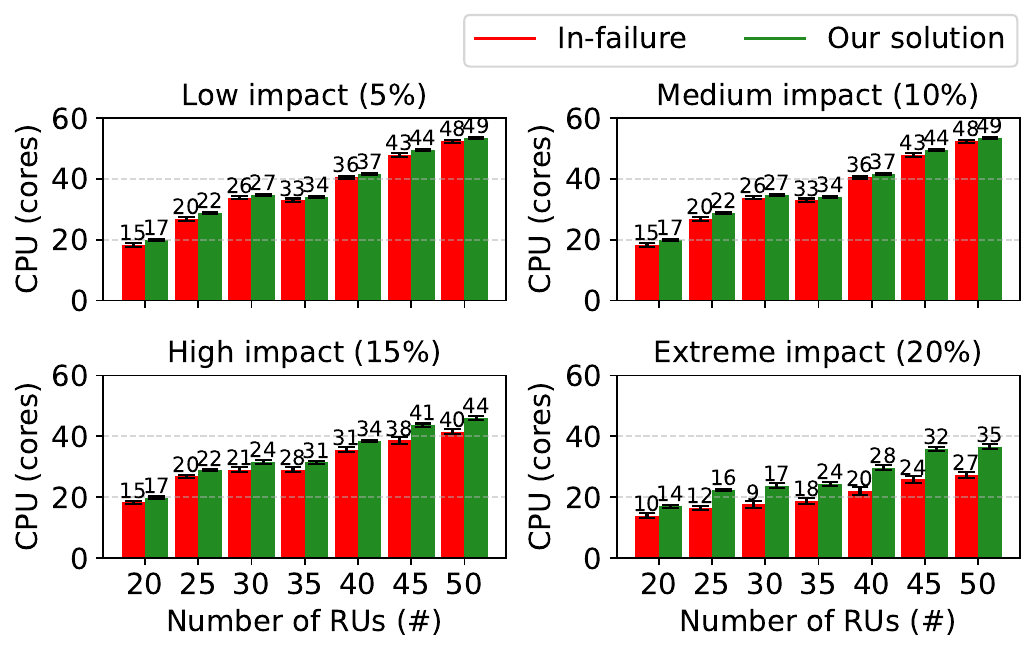}
    \vspace{-0.5em}
    \caption{CPU utilization before and after recovery. As the network restores disrupted function chains, CPU utilization increases due to the re-instantiation of \ac{CU} and \ac{DU} functions.}
    \label{fig:CPU_comparison}
    \vspace{-0.5em}
\end{figure}

In this analysis, we assess the computational cost associated with restoring service under the different failure scenarios, related to the CPU utilization required to re-instantiate \ac{CU} and \ac{DU} functions. This is essential to understanding the trade-off between achieving network resilience and allocating additional computational resources to recover disrupted function chains. Figure~\ref{fig:CPU_comparison} shows the results of our numerical evaluation, comparing the CPU utilization as the network transitions from failure to recovery. As our solution restores a larger number of function chains, especially under higher-severity scenarios, more \acp{RU} become operational, recovering service for users but also increasing overall CPU utilization. This behavior reflects the inherent trade-off between recovery performance and computational cost. The increase in CPU utilization represents the immediate cost of re-instantiating functions, but prolonged outages may pose an operational and financial strain on mobile carriers. These results highlight that recovering from disruptions requires additional (computational) resources, and practical applications of our approach may involve prioritization of critical portions of the network that serve densely populated areas or critical infrastructure. 


    \section{Conclusion}
\label{sec:conclusion}
In this work, we introduced the first resilience mechanism for disaggregated mobile networks, leveraging the adaptive placement of \ac{RAN} functions to recover from cascading failures by re-instantiating \acp{CU} and \acp{DU}. 
Our model minimizes the impact of function chain failures and their detrimental effect in the network utility, restoring service continuity in the presence of compromised infrastructure.
We evaluate our solution on real-world mobile network topologies, demonstrating its ability to mitigate failures while achieving higher performance than traditional resilience approaches. 
In future work, we plan to \1 investigate non-exact methods, including AI-based and metaheuristic approaches, to scale our resilience mechanism to larger network topologies, and \2 incorporate the criticality of \acp{RU} and the services they provide into the utility function.

\section*{Acknowledgments}
The research leading to this paper received support from the Commonwealth Cyber Initiative, an investment in the advancement of cyber R\&D, innovation, and workforce development. For more information, visit: \url{www.cyberinitiative.org}.
This work also received support from the National Science Foundation US-Ireland R\&D Partnership program under grant No. 2421362 (Resilient Networks project).
This work was supported by CAPES, by MCTIC/CGI.br/FAPESP under grant no. 2020/05127-2 (SAMURAI project), by CNPq under grant no. 306283/2025-5, by RNP/MCTIC under grant no. 01245.020548/2021-07 (Brasil 6G project), and by the OpenRAN Brazil project under grant A01245.014203/2021-14.
    
    \printbibliography

\end{document}